# Lclean: A Plausible Approach to Individual Trajectory Data Sanitization

Qilong Han[1], Dan Lu[1], Kejia Zhang[1], Xiaojiang Du[2] and Mohsen Guizani[3]

[1]College of Computer Science and Technology, Harbin Engineering University, Harbin, 150001 China
[2]Dept. of Computer and Information Sciences, Temple University, Philadelphia PA 19122, USA
[3]Dept. of Electrical and Computer Engineering, University of Idaho, Moscow, Idaho, USA

**ABSTRACT** In recent years, with the continuous development of significant data industrialization, trajectory data have more and more critical analytical value for urban construction and environmental monitoring. However, the trajectory contains a lot of personal privacy, and rashly publishing trajectory data set will cause serious privacy leakage risk. At present, the privacy protection of trajectory data mainly uses the methods of data anonymity and generalization, without considering the background knowledge of attackers and ignores the risk of adjacent location points may leak sensitive location points. In this paper, based on the above problems, combined with the location correlation of trajectory data, we proposed a plausible replacement method. Firstly, the correlation of trajectory points is proposed to classify the individual trajectories containing sensitive points. Then, according to the relevance of location points and the randomized response mechanism, a reasonable candidate set is selected to replace the sensitive points in the trajectory to satisfy the local differential privacy. Theoretical and experimental results show that the proposed method not only protects the sensitive information of individuals, but also does not affect the overall data distribution.

## I. INTRODUCTION

Over the past few years, the increasing popularity of GPS-enabled mobile devices, network communications and wearable sensors has greatly accelerated the development of location-based services (LBS) and e-commerce[1], resulting in massive amounts of trajectory data. These trajectory data refer to the set of temporal points that the user visits, and a series of Web pages that are browsed in chronological order. Using location trajectory data can help government agencies make more informed urban planning and avoid traffic jams, using web browsing records can be more customized to provide users with personalized services. However, these original trajectory data may include the user's personalized sensitive points, for example, the patient regards the hospital as a sensitive point, but the doctor may not. Privacy issues as studied in [25-28] really cause serious concern. To meet the personalized privacy needs of users, the trajectory data must be processed before release, and the sensitive points and related points should be hidden.

A great deal of work has been done on how to conduct privacy protection while mining association rules [2,3] and frequent itemsets [4]. However, they can-not be applied to the personal trajectory data directly because the trajectory data are multiple location points in chronological order. Among them, works in [5,6] adopt a suppression approach by removing the individual sensitive item, e.g., locations or web pages. The authors of [7] adopt a generalization approach to replace fine-grained data with coarse-grained data. Works in [8] adopt a permutation approach which re-arranges the ordering of items in each sensitive pattern in the released sequence. The drawback of suppression-based approaches is the data loss inflicted by deleting item in the released data, especially when sensitive patterns are frequent patterns. The generalization-based approaches can-not handle with attackers with strong background knowledge. Combined with the context, the probability of speculating on a sensitive item may increase. The permutation-based approaches break the inherent ordering in the sequential data and create "ghost" patterns that do not exist in the original database.

To the best of my knowledge, [9] is the article most similar to ours, and also protects sensitive points from opponents with strong background knowledge.Their method provides two privacy checks to decide whether to release or suppress the user's current location. The output is limited to the user's current position or suppression symbol. In other words, for the output sequence $L_1 \rightarrow L_2 \rightarrow \perp \rightarrow L_3$, the attacker can see the sensitive point is between $L_2$ and $L_3$. Combined with the context, the probability of speculating on sensitive item may increase.

Inspired by [10], this paper proposes Lclean, which applies local differential privacy to the privacy protection of sensitive points in trajectory data, which makes it impossible for attackers to infer the sensitive information of users in the original trajectory data through the published trajectory data. We assume that users set sensitive point sets themselves, and Lclean obtains these sensitive points as inputs. However, the adjacent points may also leak sensitive points, sensitive points and these sequences which may lead to the leakage of sensitive information are called sensitive regions. Our goal is to prevent any formidable attacker from inferring the user's sensitive information without affecting the data distribution too much. In this paper, we propose to replace sensitive regions with non-sensitive sequences. The replacement length depends on the correlation between sensitive points and their adjacent points. Our main contributions are as follows:

(1) We propose to apply local differential privacy to protect sensitive points in individual trajectories without affecting the usability of the overall trajectory.

(2) We propose a randomized response mechanism k-RI based on k-RR to replace the sensitive regions with other non-sensitive sequences, which do not restrict the attacker's background knowledge.

(3) We propose a novel notion, i.e., a correlation between sensitive points and their adjacent points. The strength of the correlation determines the replacement length, and if the correlation is strong, the strong correlation sequence need to be replaced. conversely, replace the sensitive point.

In the remaining of the paper, Section 2 reviews recent related works, Section 3 convers some definitions and theorems. Section 4 contains our proposed solutions. Section 5 provides experimental results. Section 6 concludes the paper and states future research directions.

## II. RELATED WORKS

Most of the location privacy protection schemes tend to avoid cryptographic primitives. The main reason is the difficulty on key management [13-15] and computational overhead. At present, the release of trajectory data can be divided into two categories. The first category is designed to handle the release of a trajectory dataset containing a large number of trajectories, each of which can be thought of as a record. The second type, [11] [12], aims to publish a single trajectory, treating each point in the trajectory as a single record. The biggest difference between the two is that the first type is the entire trajectory of the protection, while the second one is the single location. Several papers (e.g., [29-31]) have studied related privacy issues. In this article, we combine two approaches to publish a dataset that contains a large number of trajectories while protecting a user's single sensitive location.

K-anonymity based on trajectory generalization has been prevailing for its good balance of privacy protection and data availability. In [16], a (k,δ)-anonymous algorithm was proposed for trajectory dataset publication. Based on trajectory generalization and k-anonymity, this algorithm generalized every position in the trajectory to a circle with a radius of δ, and Make sure that each circle has at least k points to satisfy k-anonymity, each of which is represented by a cylinder of these circles. In [17], a data suppression technique was proposed to limit the probability of speculation of sensitive locations by assuming the attacker's background knowledge. The literature [18] considered the scenario in which users' privacy attributes were distributed with trajectory data. A (K,C)L-privacy model was proposed, where L represents the longest sub-trajectory length assumed by an attacker. The model can resist the re-identification of track data attacks and attribute attacks. However, as mentioned above, all of the techniques above for privacy distribution based on k-anonymity trajectory data set need to assume the background of the attacker, and the quantitative comparison between the different models for privacy protection can-not be made.

Differential privacy[19] was quickly applied in the privacy protection of data publishing based on fake data technology to achieve privacy protection by adding noise to the real data set. In data distribution, differential privacy can achieve different levels of privacy protection and data publishing accuracy by adjusting the privacy parameter ε. In general, the larger the value of ε, the lower the level of privacy protection and the higher the accuracy of the published dataset. The first common mechanism for implementing differential privacy is the Laplacian mechanism proposed in [20]. This mechanism mainly focuses on numeric queries. By adding random noise obeying Laplace distribution to the results of real queries. For non-numeric queries, [21] proposed an exponential mechanism, which is the second universal mechanism to achieve differential privacy.

[9] has the same purpose as our paper and protects the sensitive points of individual users while ensuring the availability of the published trajectory dataset. They provide two privacy check methods to decide whether to release or suppress the user's current location. They limit the output to real or suppressed symbols. Unfortunately, it is possible to extrapolate sensitive location information based on the suppression of symbols combined with the attacker's strong background knowledge.

## III. PRELIMINARIES

In this section, the symbols and related definitions used in this paper are given. Finally, formally defines the trajectory replacement problem.

**Location Trajectory.** A Location Trajectory is the sequence of successive spatiotemporal points produced by a particular user for a period, and it is presented as:

$$T_u = \{loc_1, loc_2, loc_3, ..., loc_n\}$$

where $loc_i$ represents the Points of Interest(POI). In the real trajectory dataset, every POI corresponds to a timestamp. In this paper, the timestamp is not taken into account when locating the sensitive region. However, when determining the candidate set, whether the replacement point meets the region reachability is judged by combining the timestamp.

**Correlation of location points.** Consider $loc_i$ as a sensitive point, if it is known that the previous location or next location can increase the probability of the attackers guess, presented as:

$$P(loc_t | loc_{t-1}) > P(loc_t) \| P(loc_t | loc_{t+1}) > P(loc_t)$$

We call it a strong correlation sequence, where $P(loc_t)$ is the probability of the attackers' random guess with the context presented as: $\max\{0.5, c(loc_t)/N\}$, where $c(loc_t)$ is the number of users who have been to $loc_t$ and N is the total number of users. $P(loc_t | loc_{t-1})$ is the probability of the attackers' infer with the previous location $loc_{t-1}$ presented as: $c(loc_{t-1}, loc_t)/c(loc_{t-1})$, where $c(loc_{t-1}, loc_t)$ is the occurrences of $loc_{t-1} \to loc_t$, and $c(loc_{t-1})$ is the occurrences of point $loc_{t-1}$.

**ε-local differential privacy.** Consider a setting where there are n records; a randomized algorithm M is $\varepsilon$-local differential privacy if and only if any two input $t$ and $t'(t, t' \in Dom(M))$, and for any possible anonymized output $O \in Range(M)$:

$$\Pr[M(t) = O] \leq e^{\varepsilon} \times \Pr[M(t') = O]$$

where the probability is taken over the randomness of $M$.

**Randomized response.** Randomized response technology [22] is the mainstream perturbation mechanism of local differential privacy. According to Kairouz et al., A gradient response technique, k-RR [23], is proposed to overcome the problem that the randomized response technique is directed to binary variables. For the case that there are k(k>2) candidate variables can also directly use randomized response. k-RR requires consistent input and output range. In this paper, In this paper, k-RI is proposed to satisfy the outputs contained in the inputs. The random response algorithm shows the output of different inputs as follows:

$$M(R) \Rightarrow \begin{cases} M(R \in Y) = \begin{cases} R & 1-P \\ R' \in Y - R & P \end{cases} \\ M(R \in X \cap Y) = R' \in Y & 1 \end{cases}$$

$R$ and $R'$ are input and output of the algorithm respectively. To make the algorithm satisfy ε-local differential privacy, we calculate $P$ by the following formula:

$$\max \frac{\Pr_{r \in R}[M(r) = R']}{\Pr_{s \in X}[M(s) = R']} = \frac{1 - P + \frac{P}{K}}{\min\left\{\frac{1}{K}, \frac{P}{K}\right\}} = \frac{1 - P + \frac{P}{K}}{\frac{P}{K}} = e^{\varepsilon}$$

$$P = \frac{K}{e^{\varepsilon} + K - 1}$$

So we adjust the randomized response mechanism to:

$$P(R'|R) = \begin{cases} \frac{1}{K} & R \notin Y \\ 1 - P + \frac{P}{K} = \frac{e^{\varepsilon}}{K - 1 + e^{\varepsilon}} & R \in Y \quad R = R' \\ P\left(1 - \frac{1}{K}\right) = \frac{1}{K - 1 + e^{\varepsilon}} & R \in Y \quad R \neq R' \end{cases}$$

If the input $R$ does not belong to the output set, then an output is selected randomly from the output set; otherwise, the probability response to the real result is $e^{\varepsilon}/(k - 1 + e^{\varepsilon})$, and any of the other $k-1$ results is $1/(k - 1 + e^{\varepsilon})$ to satisfy the local differential privacy.

**Utility loss.** The published trajectory data should be consistent with the original data distribution as much as possible, and the utility loss is defined below, which is also the most important evaluations in this experiment.

KL-divergence: For each sensitive region j to be replaced, $X = \{x_1, x_2, ..., x_n\}$ is the candidate set. $loc_p$ is the previous location of the sensitive region which is called parent node, $loc_r$ is the next location of a sensitive region which is called root node. For any $x \in X \cup S_j$, the occurrence probability of $loc_p \to x \to loc_r$ in original trajectory data is $P_j(x)$, and transformed into $Q_j(x)$ in the published trajectory. We define the kl-divergence of each sensitive region j as follows:

$$DKL_j(Q, P) = \sum_{x \in X \cup S_j} Q_j(x) \log(Q_j(x)/P_j(x))$$

Overall utility loss:

$$DKL(Q, P) = \sum_{j \in Sensitive} DKL_j(Q, P)$$

Trajectory similarity[24]: Assuming that $A = \{m_1, m_2, ..., m_i\}$ represents the original trajectory, $B = \{n_1, n_2, ..., n_i\}$ represents the processed trajectory, and the trajectory similarity between A and B is as follows:

$$TrajSim(A,B) = \frac{\max Diff(A,B) - \min Diff(A,B)}{\max Diff(A,B)}$$

Where $\min Diff(A,B)$ and $\max Diff(A,B)$ are the minimum distance and the maximum distance between A and B.

## IV. OUR PROPOSAL

In this section, we present the design of the proposed plausible approach for individual trajectory Sanitization. The overview framework is shown in Figure 4-1, that implements the end-to-end system from original trajectories and sensitive set to privacy-preserving trajectories. The original trajectory and individual sensitive points are used as the input of the whole substitution mechanism, and the sensitive points are located to determine the set of sensitive regions to be replaced. The output of the previous step is used as the input to determine the candidate set. The randomized response makes the algorithm satisfy ε local differential privacy.

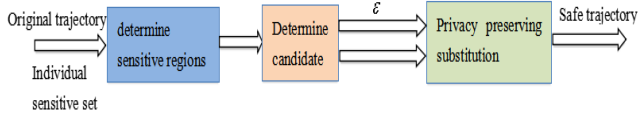

**FIGURE 1.** Framework of Lclean substitution mechanism

### A. DETERMINE SENSITIVE REGIONS

In this subsection, we present the algorithm to perform the determine sensitive regions for original trajectory data. First, we get the sensitive point index to extract the sequence containing the sensitive point. Secondly, we determine the correlation between the sensitive points and the adjacent points. If the correlation is strong, The strong correlation sequence should be replaced, on the contrary, just replace the sensitive point. Determine sensitive region algorithm (DSR) is described as follows:

| Algorithm 4.1 LSA |
|---|
| Input：original trajectory datasets $D$, the sensitive set of u $S_u$ |
| Output：set of sensitive area $SS$(including sensitive area and the former location and the latter location) |
| 1. For each $s \in S_u$ do |
| 2.    Find the former of $s$ and the latter of $s$ represented by $x_0$ and $x_1$ respectively; |
| 3.    Determine the relationship between $s$ and $x_0$, $s$ and $x_1$ according to Section 2; |
| 4.    If(Strong correlation) |
| 5.       Add the Strong correlation sequence to $SS$. |
| 6.    Else |
| 7.       Add the $s$ to $SS$; |
| 8. Return $SS$ |

Assuming that the attacker has strong background knowledge, he can do the query in the entire dataset. According to the definition in section 2, sensitive regions are obtained by calculating the correlation between each sensitive point and its adjacent points(lines 3-7).

### B. DETERMINE CANDIDATE SET

The output from the previous subsection along with the original trajectory data are used as input to this subsection. Non-sensitive regions of the same parent and child nodes as the sensitive region are found as candidate set by sub-sequence matching. This subsection considers the timestamp when determining whether a candidate set, that is to determine whether a candidate meets the spatiotemporal accessibility is judged by combining the timestamp.

$$\frac{dis(a,b)}{|t_a - t_b|} \leq \delta$$

where $\delta$ is the maximum speed of a human. If the length of the sensitive region is n, the length of candidate sequence is no more than n. Determine candidate set algorithm(DCS) is described as follows:

| Algorithm 4.2 DCS |
|---|
| Input：original trajectory dataset $D$, set of sensitive area $SS$ |
| Output：The candidate set $CS$ |
| 1. For each $ss \in SS$ : |
| 2.    Find alternate sequences do not contain sensitive information ; |
| 3.    Find alternate sequences do not have strong correlation; |
| 4.    Determine whether the alternate sequences satisfied the spatiotemporal accessibility; |
| 5.    If(satisfied) |
| 6.       Add the alternate sequence to $CS$; |
| 7. Return $CS$ |

The candidate set should satisfy three rules:

(1) have the same parent node and child node as the sensitive region(line 2);

(2) each candidate is weakly connected to the parent node and child node (line 3);

(3) A candidate value should satisfy the spatio-temporal accessibility with the parent and child nodes(line 4).

## C. PRIVACY REPLACEMENT MECHANISM

Now we get the candidate set for each sensitive region, the algorithm in this subsection applies the randomized response mechanism k-RI, which makes it impossible for the attacker to predict sensitive region with the replaced trajectory accurately. That is to say; our algorithm meets ε-local differential privacy. Above all, we assume that all k candidate sets are diverse from each other. For the candidate set containing duplicate values, the results also meet ε-local differential privacy. The privacy replacement algorithm (PRA) is described as follows:

| Algorithm 4.3 PRA |
|---|
| Input：original trajectory dataset $D$, The candidate set $CS$, $\varepsilon$ |
| Output：noisy trajectory dataset $D_r$ |
| 1. $\varepsilon = \varepsilon / |CS|$; |
| 2. For each $S_j \in S_u$ |
| 3. For each $seq_i \in CS_j$: |
| 4. Algorithm A determine the relationship between $seq_i$ and the former location and the **correlation** between $seq_i$ and the latter location. |
| 5. If(weak correlation) |
| 6. $K = |CS_j|$, find a $seq_i \in CS_j$ as the input, according to the formula 4-1, the possibility of output $seq_i$ is $e^\varepsilon / (k - 1 + e^\varepsilon)$, and the possibility of picking one randomly from the rest is $1 / (k - 1 + e^\varepsilon)$. |
| 7. Replace the sensitive area with the output. |
| 8. Return $D_r$ |

This subsection implements the substitution work for each sensitive region. First, we divide the privacy budget. The candidate set:
$$CS = \{\{SS_i, seq_1, seq_2, ..., seq_{|n|}\}...\{SS_j, seq_1, seq_2, ..., seq_{|m|}\}\}$$

The size of $CS$ is the number of sensitive regions to be replaced. Therefore, the privacy budget for each sensitive region is $\varepsilon / |CS|$ to ensure that the entire trajectory meets ε-local differential privacy. This equal division of the privacy budget is called the average privacy replacement(BR). Each candidate is evaluated for correlation before replacement to ensure that it is weakly correlation after the replacement(line3-5). Line6 implements local differential privacy replacement, and the output of randomly response mechanism is used as the final replacement value.

## D. PROPORTIONAL PRIVACY REPLACEMENT

In the previous subsection, to guarantee the local differential privacy of the entire trajectory, we first assign an average privacy budget when replacing each sensitive region. In this section we give a ratio-based allocation of the privacy budget(RatioR).

$$\varepsilon_j = \varepsilon \times |CS_j| / \sum_{i=0}^{|CS|} |CS_i|$$

Where $|CS_j|$ is the candidate set size for the j sensitive region. In the real trajectory, the size of candidate set for each sensitive region to be replaced varies from region to region. In this section, the region to be replaced with fewer candidates is allocated smaller privacy budget to ensure a stronger degree of privacy protection and more privacy budget for larger candidates.

## V. PERFORMANCE EVALUATION

To simulate the location trajectory anonymization we have used the Gowalla check-in dataset. This dataset was collected by almost 200000 users, and it contains 650000 trajectories. To verify the utility of our method, we first define a baseline method [8] in which the sensitive points (sequences) are published in the form of. The utility loss function defined in Section 3 is taken as our experimental evaluation target: KL-divergence. We compare our average privacy replacement (BR) and the ratio-based privacy replacement (RatioR) with the baseline method under different privacy budgets.

### A. KL-DIVERGENCE

Based on the KL-divergence metric in Section 3, the error between the replaced trajectory data and the original trajectory data is quantified. In this paper, we replace and add noise to multiple sensitive regions, the practicality measure in this subsection is evaluated separately for each sensitive region. For any sensitive region to be replaced, the candidate set is $CS_j$, the probability of outputting any one of the candidate set is:

$$\Pr_{c_i \in CS_j}(c_i) = \frac{count(c_i)}{|CS_j|} \times \frac{e^\varepsilon}{|CS_j| - 1 + e^\varepsilon} + \frac{1}{|CS_j| - 1} \times \frac{1}{|CS_j| - 1 + e^\varepsilon}$$

We adjust the KL-divergence formula to:

$$\sum_{i=0, c_i \in CS_j}^{|CS_j|} \Pr(c_i) \times \left( \frac{count(loc_p \to c_j \to loc_l) + 1}{count(loc_p \to * \to loc_l)} \log \left( \frac{count(loc_p \to c_j \to loc_l) + 1}{count(loc_p \to c_j \to loc_l)} \right) \right)$$

Where $count(loc_p \to c_i \to loc_l)$ represents the number of occurrences of $loc_p \to c_i \to loc_l$ in the trajectory dataset, $count(loc_p \to * \to loc_l)$ represents the number of locations that the previous location is $loc_p$ and the next location is $loc_l$. Figure 5-1 shows the utility loss compared with the privacy replacement method BR and RatioR.

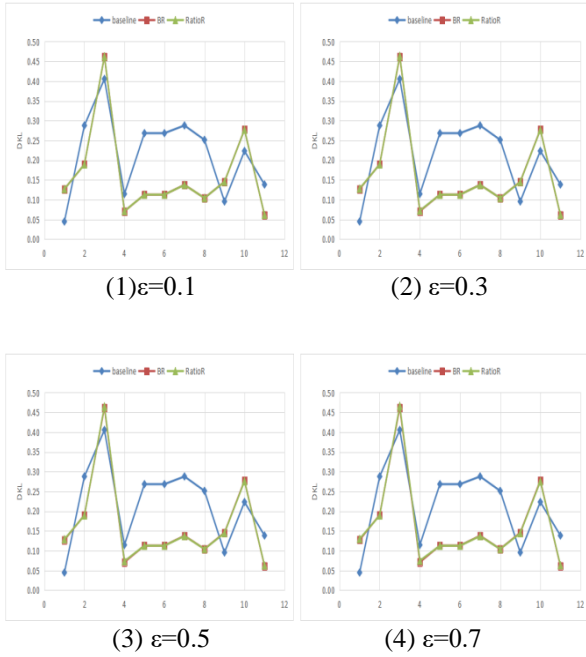

(1) ε=0.1    (2) ε=0.3

(3) ε=0.5    (4) ε=0.7

**FIGURE 2.** Utility loss comparison

The abscissa is the number of the sensitive region to be replaced, and the ordinate is the utility loss of DKL. It can be seen from the figure that the loss of utility of our average privacy replacement method (BR) and the ratio-based privacy replacement method (RatioR) is lower than that of the baseline method. For the point of 3 and 10, the utility loss is slightly higher than the baseline method. This is a result of having a too small candidate set, which will lead to a great fluctuation with a certain value changes. Also, we can see from the experiment that RatioR increases the degree of protection for each sensitive region and thus reduces its practicability, the possible loss $DKL_j$ increases slightly.

The utility loss of the entire trajectory is shown in Figure 5-2 below:

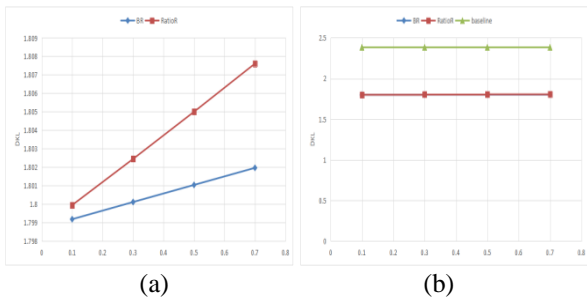

(a)    (b)

**FIGURE 3.** Utility loss of the whole trajectory

As shown in the figure, the abscissa is the value of the privacy budget. It can be seen from Figure a that as the privacy budget increases, the practicality of the overall trajectory data increases slightly. From Figure b, we can see that the utility loss of our two methods is significantly lower than the baseline method. The experimental results show that our average privacy replacement method (BR) and the ratio-based privacy replacement method (RatioR) meet ε-local differential privacy while maintaining better practicality.

### B. TRAJECTORY SIMILARITY

Based on the trajectory similarity metric in Section 3, the similarity between the replaced trajectory data and the original trajectory data is quantified. We change the privacy parameter ε and conduct multiple experiments to get the average. Figure 5-3 shows the trajectory similarity results with the privacy replacement method BR and RatioR.

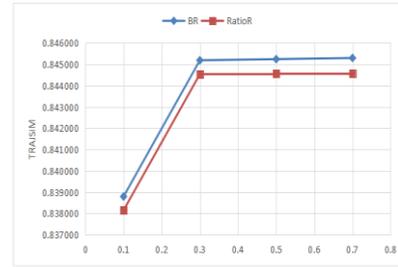

**FIGURE 4.** Trajectory similarity under different ε

As shown in the figure, the abscissa is the value of the privacy budget. It can be seen that as the privacy budget increases, the similarity value increases. We can see that when ε increases from 0.1 to 0.3, the trajectory similarity increases greatly, but there was no significant change from 0.3 to 0.7. The experimental results show that the released trajectory under ratio-based privacy replacement method (RatioR) is more similar to the original trajectory.

## VI. CONCLUSION

We presented Lclean, a system that uses substitution as a mechanism to protect sensitive points of an individual. Instead of random substitution of sensitive regions, which can degrade the utility of the overall datasets, we perform a privacy substitution mechanism with a randomized response. We employ an average privacy substitution and a ratio-based privacy substitution to meet the ε-local differential privacy while maintaining the consistency of release trajectory. Through experimentation on real-life check-in trajectory datasets, we demonstrated that our two privacy substitution strategies could indeed be used for preserving the utility of processed data while achieving ε-local differential privacy. In future, we have the plan to work on multi-user trajectory privacy protection, considering the connection between different users' sensitive points and make it difficult to figure out person sensitive information.